\input psfig

\documentstyle{article}

\begin{document}
\large

\Large
\centerline
{\bf Partitioning of a Polymer Chain between Two Confining Cavities:}
\centerline
{\bf The Roles of Excluded Volume and Finite Size Conduits}

\vspace{0.5 cm}
\large
\centerline{Stefan Tsonchev and Rob D. Coalson}
\centerline{Department of Chemistry}
\centerline{University of Pittsburgh}
\centerline{Pittsburgh, PA 15260}

\vspace{1. cm}
\normalsize

\underline{Abstract:}
Lattice-field calculations are performed on a Gaussian polymer 
chain confined 
to move within the region defined by two fused spheres. The results
of the calculations are in accord with recent experimental measurements and
computer simulations, and suggest that current theoretical understanding of
polymer partitioning phenomena is not adequate when excluded 
volume interactions between the monomers are present. It is also shown that
the notion of ground state dominance can fail even in the large monomer
limit.

PACS numbers: 61.82.Pv, 83.70.Hq, 05.90.+m

\vspace{0.8 cm}
\large
A number of technologically important processes such as gel electrophoresis,
size exclusion chromatography and membrane separation \cite{genref}
depend on the partitioning of a polymer chain between two or more 
confining cavities connected by small conduits.  The goal is
to utilize the dependence of the partitioning on molecular properties
such as polymer length or, in the case of polyelectrolytes, electrical charge
and electrolyte composition,
to selectively separate polymer chains, e.g. according to
their molecular weight.

Current theoretical understanding of polymer
partitioning via the principles of equilibrium statistical
mechanics
\cite{cas,muth} 
is based on the following
notions.  The Helmholtz free energy $A$ of a polymer chain of length
$M$ is estimated as $\beta A \cong M (b/R)^{1/\nu}$, where
$\beta^{-1} = kT$, $b$ is the Kuhn length of the polymer,
$R$ is the characteristic linear dimension of the cavity, and $\nu = 1/2,3/5$
for chains without and with excluded volume, respectively.  The
argument then goes that if two cavities with different sizes are
connected by a narrow conduit (such that thermal equilibrium
is established between them), the polymer chain will partition
itself such that the ratio of the number of monomers in each cavity (designated
$1$ and $2$) is:

\begin{equation}
K \equiv \frac{M_1}{M_2} = \exp \left[ -\beta (A_1 - A_2) \right]
\cong  \exp \left[- (const)M  \left(\frac{1}{R_1^{1/\nu}} -
\frac{1}{R_2^{1/\nu}} \right) \right ]  \;  ,
\label{eq:muth1}
\end{equation}

where $const$ is a constant of order unity. This leads immediately to 
the conclusion that the polymer prefers
to occupy the larger cavity (say, cavity $1$), and in particular
that $\ln{K}$ grows linearly with the polymer chain length.

There are several approximations involved in arriving at 
Eq. (\ref{eq:muth1}).  The estimation of $\beta A_{1,2}$ assumes
that ground state dominance of the Green's function
governing the distribution of monomers associated with the
polymer \cite{dgde} applies.
Naively, this requires that the polymer chain
in cavity $1,2$ be sufficiently long that $\beta A_{1,2} \gg 1$
(but see below).  Furthermore, incorporating the effects of
excluded volume by modifying the
exponent $\nu$ from the value of $1/2$ (valid for a simple
Gaussian chain \cite{dgde}) to $3/5$ requires the application of rough
scaling arguments.  

Eq. (\ref{eq:muth1}) seems intuitively
reasonable when excluded volume effects are suppressed,
because the polymer chain gains entropy when it passes from
the smaller to the larger sphere.  However, when excluded
volume is included, intuition suggests that there will be
a point beyond which putting more monomers in the larger sphere
will cost, rather than lower, free energy, due to inter-monomer
repulsion.  Indeed, recent experimental partition coefficient
measurements indicate the existence of such a saturation effect.
In these experiments \cite{asher} a large (100 nm) spherical cavity was
etched into a gel polymer network.  
Pockets in the gel network, typically 5--10 nm in linear
dimension, play the
role of the ``small" cavity in the above arguments.
${\ln}K$ was indeed found to grow nearly linearly with $M$ for small
$M$, but a slower (sublinear) growth rate was observed as
$M$ was increased.

Recent Langevin dynamics simulations on a model of this gel/cavity system show
similar effects \cite{chern}.  The simulations also indicate a monotonic
increase in ${\ln}K$ with $M$, followed by a turnover regime
(as excluded volume effects become significant).  At large
$M$, ${\ln}K$ was found in the simulations to {\it decrease}
with $M$.  Clearly, intuition, experiments and simulations
all suggest that Eq. (\ref{eq:muth1}) needs to be modified when
excluded volume effects become significant.

In this letter we consider the equilibrium partitioning ratio of
a Gaussian polymer chain
in a container comprised of two spheres of different radii connected
by a small aperture (cf. Fig. 1).
When excluded volume effects are
neglected, numerically exact calculations of the
monomer density in either sphere
can be performed by solving an appropriate three-dimensional (3D)
 Schr\"odinger Equation.
We have carried out such calculations using simple real-space
lattice methodology.  When excluded volume is included,
a mean-field solution
can be obtained by solving the same Schr\"odinger Equation with
a modified effective potential that depends self-consistently
on the monomer density.  We have also carried out calculations
of this type.  These shed light on both the no-excluded
volume limit and the effect of significant excluded volume.
We also study the effect of aperture size on the results.
The limit of an infinitesimal aperture, which allows thermal
contact and material transfer without significantly altering the
topologies of the two individual spheres, is conceptually important
but not necessarily experimentally realistic.  Finite size apertures
are found to modify the behavior of the system significantly in
some respects.

For the case of a single polymer represented via the Gaussian
Chain Model, the equilibrium properties are completely determined
in the absence of intermonomer excluded volume interactions
by the eigenfunctions and eigenvalues of the following 3D
Schr\"odinger Equation \cite{dgde}:

\begin{equation}
\left[
- \frac{b^2}{6} \nabla^2 + V (\vec r) /kT  \right] \psi (\vec r) =
E \psi (\vec r) \; .
\label{eq:se1}
\end{equation}

Here $V(\vec{r})$ is an externally supplied potential energy function
experienced by each monomer in the polymer.
In the present case $V$ is zero inside the container and infinite
at the container walls.  Thus, the effective quantum mechanical problem
is that of a particle in the box indicated in Fig. 1.  Determination
of the eigenvalues/vectors for this problem must be done numerically.
Although the system shown in Fig. 1 has cylindrical symmetry, we
have chosen to develop a numerical method for solving the Schr\"odinger
Equation which is valid for arbitrary 3D bound state problems.
Specifically, we use a finite-difference position space representation,
in which the wavefunction is described on a cubic real
space lattice \cite{tson}.
In this representation the
potential energy matrix is diagonal (with diagonal
value equal to the value of the potential at a given lattice point)
and the kinetic energy entails off-diagonal coupling between
nearest neighbors, as prescribed by simple symmetric finite-differencing
of the Laplacian.  Since the overall Hamiltonian matrix is sparse,
low lying  energy eigenfunctions and eigenvectors can be computed
efficiently via a Lanczos algorithm \cite{lan}.

\begin{figure}[!]
\psfig{file=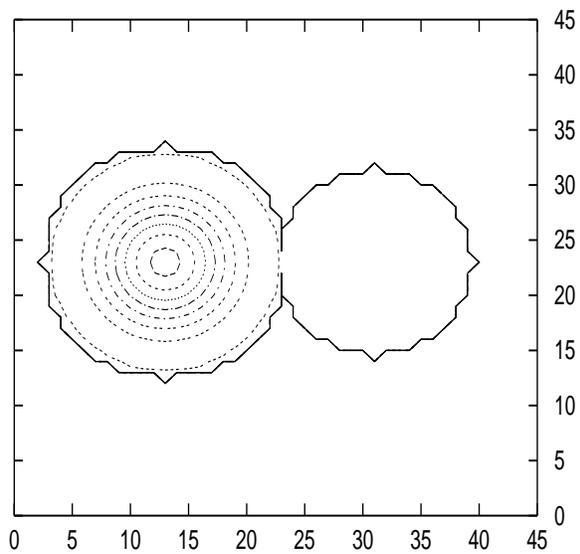,width=370pt,height=370pt,angle=270}
\vspace{-0.8in}
\caption{Slice through a plane containing the symmetry
axis of two spheres with radii $R_1 = 1.0$ (left) and
$R_2 = 0.8$ (right) connected by a narrow (``minimally-fused") aperture.
(Imperfections in the shape of the confining container are
lattice artifacts.)
Solid line
shows outline of the confining system.
Also shown, via dashed lines,
is a contour plot of
$\psi_{0}^{2}(\vec{r})$ in this plane for the case of $\lambda=0.0$
(no excluded volume).  Note that this function is completely
confined to the large sphere.}
\end{figure}

Given the eigenvalues $E_j$ and unit-normalized eigenfunctions
$\psi_j$ of the Schr\"odinger Equation 
(\ref{eq:se1}), the monomer density can be synthesized as:

\begin{equation}
\rho(\vec{r})=\frac{\sum_{j=0}^{\infty}
\sum_{k=0}^{\infty}A_{j}A_{k}\psi_{j} (\vec r)
\psi_{k} (\vec r) f(M;E_{j},E_{k})}
{\sum_{j=0}^{\infty}A_{j}^{2}e^{-ME_{j}}} \; ,
\label{eq:rho}
\end{equation}

where $A_j=\int d\vec r \psi_j (\vec r)$,
and

\begin{equation}
f=\left\{ \begin{array}{cc} 
\frac{e^{-ME_{j}}-e^{-ME_{k}}}{E_{k}-E_{j}}\,\,\,\,\,; \,for \,\,E_{j} \neq 
E_{k}&\\
Me^{-ME_{j}} \,\,\,\,\,\,\,\,\,\,\,\,\,\,\,; \,for \,\,E_{j} = E_{k}&
\end{array} \right. \; .
\end{equation}

The $f$ factors suppress high-lying excited states (more quickly
as $M$ increases), so that the sums in Eq. (\ref{eq:rho}) can be
truncated at a finite value using the information obtained via
diagonalization of
a finite-dimensional Hamiltonian matrix associated with a
particular lattice size.   

Since the energy levels $E_j$ are independent of $M$ when
the potential energy function $V(\vec r)$ is externally
prescribed, the condition $M(E_1 - E_0) \gg 1$ is guaranteed
for long polymer chains.
In this limit ground state dominance occurs, i.e.,
only the ground state of the Hamiltonian in Eq. (\ref{eq:se1}) 
affects the thermodynamics of the system, and in particular
$\rho (\vec r) \rightarrow M \psi_0^2 (\vec r)$.

Results of calculations for the parameters $R_1 = 1.0$, $R_2 = 0.8$ and
$b = 0.2$ for a polymer confined to the volume of two spheres sharing one 
common point on the lattice are shown in Figs. 1--2. (A lattice 
of 44 points per side was used throughout.)  Fig. 1 shows 
$\psi_0^2 (\vec r)$, i.e., the square of the
ground state eigenfunction, for this ``minimally-fused"
configuration. Note that 
$\psi_0^2(\vec{r})$ lies entirely in sphere $1$.  The first excited state
(not shown) is similar in shape, but is entirely confined
to sphere $2$.
In fact, due to the impenetrable walls of the container
and the tiny contact region, the eigenstates divide into sets,
one set describing a particle in spherical cavity
$1$ and a second set associated with
sphere $2$.  Because of this character,
the monomer densities in spheres 1 and 2 are easily calculated (cf. 
Eq. (\ref{eq:rho})), and the partition coefficient is given by:

\begin{equation}
\frac{M_1}{M_2} =
\frac{\sum_{j=0}^\infty {A_j^{(1)}}^2 e^{-M E_j^{(1)}}}
{\sum_{j=0}^\infty {A_j^{(2)}}^2 e^{-M E_j^{(2)}}}
\rightarrow
\frac{{A_0^{(1)}}^2}
{{A_0^{(1)}}^2}
 e^{-M \left[ E_0^{(1)} - E_0^{(2)} \right]} \; ,
\label{eq:pc1}
\end{equation}

where the superscript $(1)$ denotes states localized in sphere $1$, 
analogously for $(2)$, and the sums in numerator and denominator are
over all states localized in spheres $1$ and $2$, respectively.
The large M limit is indicated by the arrow.



Noting that the ground state energy of a particle in a spherical
box of radius $R$ is $E_0 = \pi^2 b^2 / 6R^2$, we see that
Eq. (\ref{eq:pc1}) is in perfect harmony with the estimation of
Muthukumar and Baumg\"artner, Eq. (\ref{eq:muth1}) (with $\nu = 1/2$).
The converged ${\ln}K$ vs. $M$ curve, shown in Fig. 2, confirms
the rapid onset of the limiting behavior prescribed by Eq. 
(\ref{eq:pc1}).

\begin{figure}[!]
\psfig{file=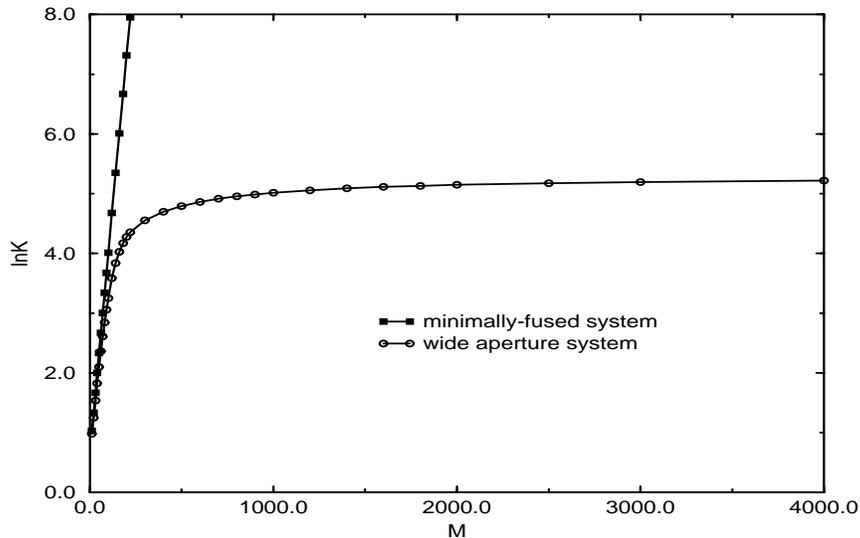,width=370pt,height=240pt,angle=270}
\caption{ $\ln{K}$ vs. $M$ for $\lambda=0.0$.  Squares
depict result for the minimally-fused aperture system;
circles depict the analogous result for the wide aperture
system illustrated in Fig. 3.}
\end{figure}

It is interesting to consider the effect of a wider aperture
on polymer partitioning between the same two spheres considered
above.  The specific ``wide-aperture" case we will study is
shown in Fig. 3 with the two spheres moved inside each other by one 
more lattice spacing.  We expect that the eigenstates of a Hamiltonian
with this confining potential will not be completely localized
in one sphere or the other.  This expectation is born out by
explicit numerical calculation.  $\psi_0^2(\vec{r})$ 
is shown in Fig. 3 for the  values of $R_1$, $R_2$ and $b$
noted above (which are used in all calculations presented below).
Note in particular that the ground state
of this fused two-sphere system has some ``leakage" into the
the smaller sphere.  This means that in the large polymer chain
limit, where the monomer density is determined solely by $\psi_0^2(\vec{r})$,
$M_1/M_2$ saturates at a finite value, as shown in Fig. 2,
rather than tending to infinity as it does in the limit of
an infinitesimal aperture.

\begin{figure}[!]
\psfig{file=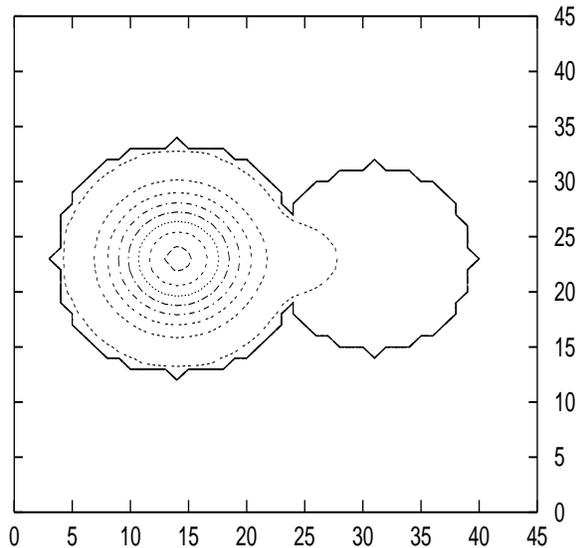,width=370pt,height=370pt,angle=270}
\vspace{-0.8in}
\caption{Slice through a plane containing the symmetry
axis of two spheres with radii $R_1 = 1.0$ (left) and
$R_2 = 0.8$ (right) connected by a wide aperture.  Solid line
shows outline of the confining system.
Also shown, via dashed lines,
is a contour plot of
$\psi_{0}^{2}(\vec{r})$ in this plane for the case of $\lambda=0.0$
(no excluded volume).
Note the leakage of this function into
the small sphere.}
\end{figure}

The situation becomes more complicated 
when excluded volume between monomers is incorporated into
the model.  Adopting a mean-field description of excluded volume
effects leads to the following modification of the quantum-mechanical
isomorphism utilized above.  Namely, the relevant effective 
Schr\"odinger Equation becomes \cite{dgde},

\begin{equation}
\left[
- \frac{b^2}{6} \nabla^2 + \lambda \rho(\vec r)  \right] \psi (\vec r) =
E \psi (\vec r) \; .
\label{eq:se2}
\end{equation}

Here $\lambda > 0$ is the excluded volume parameter, which
has the dimensions of volume and can be approximately identified
with the cube of the monomer radius.  Since the effective
potential is now a functional of the eigenfunctions of the Hamiltonian
operator, a nonlinear Schr\"odinger Equation must be solved.  We do this
by an iterative process in which an initial density profile is
``guessed" (e.g., the density corresponding to the $\lambda =0$
limit), the Schr\"odinger Eq. (\ref{eq:se2}) is solved using the
discretized real-space lattice/Lanczos procedure described
above, the eigenfunctions
and eigenvalues obtained from this calculation are used to 
compute a new density (via Eq. (\ref{eq:rho}) above), and the
cycle is repeated until self-consistency is achieved.  In practice,
since this Schr\"odinger Equation is highly nonlinear, care has
been taken in order to prevent the onset of numerical instabilities \cite{tson}.

We considered the same minimally fused system studied above
in the $\lambda = 0$ limit for the case $\lambda = 0.001$.
The dependence of the natural log of the
partition coefficient on polymer length $M$ is shown in Fig. 4.
Note that for small $M$ the effects of excluded volume  
are negligible, i.e., ${\ln}K$ grows nearly linearly
with $M$, in accord with the $\lambda=0.0$
case.  However, as the polymer length increases, ${\ln}K$ increases
{\it sub}linearly with $M$.  As $M$ increases further, 
${\ln}K$ reaches a maximum and then begins to {\it de}crease.
At very large $M$ an asymptotic value, bound from below by the
natural log of the
volume ratio of the two spheres, is obtained.  All these
features are in agreement with the expectations expressed
at the outset, and also consistent with experiments and
simulations on a spherical cavity embedded in a gel 
network, but are not contained in Eq. (1).

\begin{figure}[!]
\psfig{file=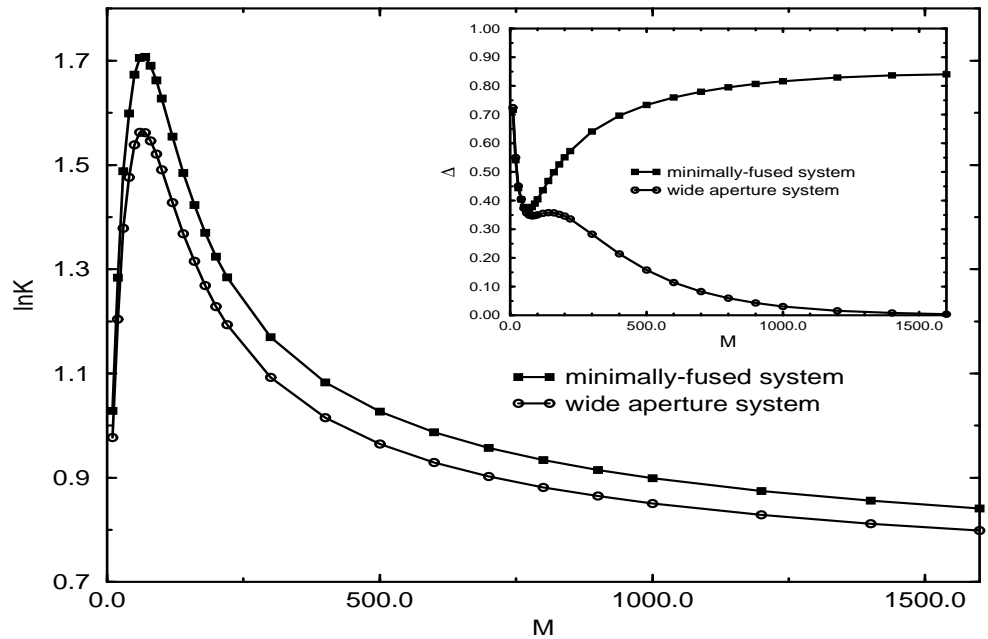,width=370pt,height=240pt,angle=270}
\caption{Main panel shows $\ln{K}$ vs. $M$ for $\lambda=0.001$.
Squares show results for the minimally-fused aperture;
circles show analogous results for the wide aperture case.
The inset shows the number $\Delta{\equiv}\exp[-M(E_{1}-E_{0})]$ as a 
function of $M$.} 
\end{figure}

An important observation is that for the  mean-field
excluded volume model considered here, {\it ground state
dominance in the large chain-length limit
does not necessarily occur}.  In fact, in the present minimally-fused
spheres example, the notion of ground state dominance fails
manifestly. In the inset to Fig. 4 we plot 
$\Delta{\equiv}\exp[-M(E_{1}-E_{0})]$ vs. $M$,  
as an indicator of ground state dominance ($\Delta \rightarrow 0$
as $M \rightarrow \infty$).
It can be seen that ground state dominance does not occur 
in the minimally-fused spheres example. 
Because the energy levels $E_j$ depend on $M$ in this
nonlinear Schr\"odinger Equation, it is not guaranteed 
that $M (E_1 - E_0) \gg 1$ as $M \rightarrow \infty$. 
Explicit calculation shows that the gap between $E_0$ and
$E_1$ narrows in such a way that this condition never arises. Even
in the large $M$ limit the first excited state must be retained in the
calculation in order to obtain the correct large $M$ limit of the 
partition coefficient $K$.

We have also considered the effect of a wide aperture on
the Gaussian polymer with excluded volume.  As indicated in
the inset to Fig. 4, in this case ground state dominance {\it does} obtain
in the large $M$ limit, consistent with the behavior of 
a single cavity system (e.g., a single sphere or ellipsoid).
The ${\ln}K$ vs. $M$ curve, shown in the main panel of Fig. 4, has the same 
generic shape as for the minimally fused case.

{\it In summary}, we have used a finite-difference representation
of the 3D Schr\"odinger Equation to compute the equilibrium
partition coefficient for a single Gaussian polymer chain
in a system consisting of two spheres of unequal sizes connected
by a narrow conduit.  If excluded volume effects are neglected,
this approach provides numerically exact solutions for the
Gaussian chain model.  Such calculations show that
for a very narrow conduit, the estimation previously provided
by Muthukumar and Baumg\"autner \cite{muth} is accurate,
while for a wider conduit
the partition coefficient saturates at a large but finite value.
When excluded volume is included at the mean field
level the situation changes considerably.  For small $M$ the
system shows nearly linear growth of ${\ln}K$ with $M$ (essentially,
excluded volume corrections are unimportant), while for larger
$M$, this curve achieves a maximum and then falls to an
asymptote at infinite $M$ which reflects the volume ratio
of the two spheres. The possibility of failure of the notion of ground state
dominance in this system emphasizes the importance of carefully considering
the effects of excited state contributions to the relevant Green's
functions, even in the large chain length limit.
Extension of this analysis
to the case of a charged polymer chain in the presence of
electrolyte solution
will be considered in subsequent work.

\vspace{1.5 cm}
\underline{Acknowledgements:} We thank Anthony Duncan for many
valuable discussions.  This work was partially supported
by National Science Foundation grant No. CHE-9633561.

\end{document}